\documentclass[%
 reprint,
nofootinbib,
 amsmath,amssymb,
 bibtex,
 aps,
]{revtex4-2}

\usepackage{xcolor}

\usepackage[normalem]{ulem}

\usepackage[export]{adjustbox}

\usepackage{graphicx}
\usepackage{dcolumn}
\usepackage{bm}
\usepackage{amsmath, amssymb}
\usepackage{graphicx}

\usepackage{multirow}

\usepackage{tikz}
\usepackage{tkz-euclide}
\usetikzlibrary{shapes.misc}
\usetikzlibrary{decorations.pathmorphing}	
\tikzset{
    v/.style={decorate, decoration={snake, segment length=3mm, amplitude=0.75mm}, draw},
    f/.style={draw=black, postaction={decorate},
        decoration={markings,mark=at position .6 with {\arrow[very thick]{latex}}}},
    fb/.style={draw=black, postaction={decorate},
        decoration={markings,mark=at position .4 with {\arrowreversed[very thick]{latex}}}},
    fnar/.style={draw=black},
    g/.style={decorate, draw=black,
        decoration={coil,amplitude=3pt, segment length=3.5pt}},
    s/.style={dashed,draw=black, postaction={decorate},
        decoration={markings,mark=at position .55 with {\arrow[very thick]{latex}}}},
    sb/.style={dashed,draw=black, postaction={decorate},
        decoration={markings,mark=at position .55 with {\arrowreversed[draw=black,very thick]{latex}}}},
    snar/.style={dashed,draw=black,line width =1.25pt},
    cross/.style={cross out, draw=black, minimum size=2*(#1-\pgflinewidth), inner sep=0pt, outer sep=0pt},
cross/.default={3pt},
}

\def\be{\begin{equation}}
\def\ee{\end{equation}}

\newcommand{\braket}[1]{\left\langle #1 \right\rangle}
\newcommand{\sbraket}[1]{\left[ #1 \right]}
\newcommand{\asbraket}[3]{\left\langle #1 \vert #2 \vert #3 \right]}
\newcommand{\sabraket}[3]{\left[ #1 | #2 | #3 \right\rangle}
\newcommand{\bra}[1]{ \left\langle #1 \right|}
\newcommand{\sbra}[1]{ \left[ #1 \right|}
\newcommand{\ket}[1]{ \left| #1 \right\rangle}
\newcommand{\sket}[1]{ \left| #1 \right]}
\newcommand{\order}[1]{ \mathcal{O}\left( #1 \right)}

\newcommand{\al}[1]{\begin{align}\begin{aligned} #1 \end{aligned}\end{align}}

\begin{document}

\preprint{APS/123-QED}

\title{An on-shell approach to neutrino oscillations}
\author{Gustavo F. S. Alves}%
 \email{gustavo.figueiredo.alves@usp.br}
\author{Enrico Bertuzzo}
\email{bertuzzo@if.usp.br}
 \author{Gabriel M. Salla}%
 \email{gabriel.massoni.salla@usp.br}

\affiliation{%
 Instituto de F\'isica, Universidade de S\~ao Paulo, C.P. 66.318, 05315-970 S\~ao Paulo, Brazil
}%




\date{\today}

\begin{abstract}
In the usual quantum field theoretical approach, neutrino oscillations are studied diagonalising either the mass or matter Hamiltonians. In this paper we analyse the problem from an on-shell amplitude perspective, where Lagrangians or Hamiltonians are not available. We start by studying in detail how flavor enters in the amplitudes and how the PMNS matrix emerges. We then analyse the elastic amplitude of two neutrinos and two charged leptons that induce matter effects and propose a strategy to obtain the known results of the standard oscillation theory without Hamiltonians. Finally, we extend the previously proposed procedure and use the most general elastic 4-point amplitude to study beyond the Standard Model effects on oscillations.


\end{abstract}

\maketitle


\section{\label{sec:intro}Introduction}

Many decades after their discovery, neutrino oscillations are still among the most interesting and subtle observed phenomena, and one of the few clearly established evidences of Beyond the Standard Model (BSM) physics~\cite{Giunti:1053706,barger2012physics}. Two complementary approaches have been used to describe oscillations: a Quantum Mechanical one, in which the neutrino wave packet propagates between the production and detection regions, and a Quantum Field Theory (QFT) one, in which production, propagation and detection are considered as a unique process. For a review and a comparison between the two formalisms we refer the reader to~\cite{Akhmedov:2010ms} and references therein.
Besides the oscillations in vacuum, interactions with matter, the Mikheyev-Smirnov-Wolfenstein (MSW) effect \cite{Wolfenstein:1977ue,Mikheyev:1985zog,Barger:1980tf}, also play a major role in neutrino physics. The usual description of this effect strongly relies on the computation of the interacting Hamiltonian, which in turn involves further diagonalization of the neutrino states (see for instance \cite{Linder:2005fc}).

In this paper we are interested in approaching the oscillation phenomenon from the perspective of on-shell methods~\cite{Dixon:2013uaa,Arkani-Hamed:2017jhn,Chung:2018kqs}.~\footnote{See Ref.~\cite{Li:2021tsq} for an application of on-shell techniques to the determination of the operator up to dimension 9 in the SMEFT with light sterile neutrinos.} Such methods bypass completely the need for quantum fields, Lagrangians and Hamiltonians, constructing scattering amplitudes directly in terms of physical states and “fundamental” 3-point amplitudes. Moreover, the on-shell program has a remarkable property regarding amplitudes in Effective Field Theories (EFTs)~\cite{Christensen:2018zcq,Shadmi:2018xan,Ma:2019gtx,Aoude:2019tzn,Durieux:2019eor,Bachu:2019ehv,Durieux:2019siw,Gu:2020thj}: once the particle content of an $n$-point amplitude is chosen, on-shell techniques allow to enumerate all possible kinematic structures permitted by the little group, independently of the order in the EFT at which they are generated for the first time. Scattering amplitudes are thus a powerful tool to include all-order beyond the Standard Model (BSM) effects without having to worry about operators basis and fields redefinitions. The main questions we want to address in this paper are: (i) how may flavor be included in on-shell amplitudes and how do we recover the unitarity of the Pontecorvo–Maki–Nakagawa–Sakata (PMNS) matrix without quantum field redefinitions? (ii) how may we obtain the MSW effect in the absence of any matter Hamiltonian?


This paper is organised as follows. In Sec.~\ref{sec:PMNS} we study the 3-point amplitude of one neutrino, one charged lepton and a $W$ boson. In particular, in Sec.~\ref{sec:3_point_amp} we study its high-energy behaviour, making contact to the usual operator language. Then, in Sec.~\ref{sec:Emergence}, we discuss how to implement flavor quantum number at the level of amplitudes and show how the PMNS matrix naturally appears in this framework. Sec.~\ref{sec:oscillations} is instead dedicated to the computation of matter effects: in Sec.~\ref{sec:potential} we present a way to compute the matter potential, while in Sec.~\ref{sec:propagation} we propose a strategy allowing to obtain the modified dispersion relations for neutrinos typical of propagation in matter. Finally, in Sec.~\ref{sec:BSM} we compute the matter effects in the presence of general effective interactions. We conclude in Sec.~\ref{sec:conclusions}.

\section{\label{sec:PMNS} The PMNS matrix from the $\nu \bar{e} W$ amplitude}

\subsection{\label{sec:3_point_amp} The 3-point amplitude} 

Our starting point is the 3-particle massive amplitude involving one neutrino, one charged lepton and one $W$ boson. Its structure in terms of spinor helicity variables is~\cite{Durieux:2019eor}
\al{\label{eq:3-point_noflavor}
\mathcal{A}\left[ 1_\nu 2_{\bar{e}} 3_W \right] & =  \frac{y_L}{M} \braket{\bm{13}} \braket{\bm{23}} + \frac{g_L}{m_W} \braket{\bm{13}} \sbraket{\bm{23}} \\
& + \frac{g_R}{m_W} \sbraket{\bm{13}} \braket{\bm{23}} + \frac{y_R}{M} \sbraket{\bm{13}} \sbraket{\bm{23}} ,
}
where we have adopted the bold notation of Ref. \cite{Arkani-Hamed:2017jhn} for massive spinors (see App. \ref{app:conventions_spinor_helicity}).
Note that the second and third terms are suppressed by the $W$ mass $m_W$, while the first and last term are suppressed by some large mass scale $M \gg m_W$. The reason behind this resides in the high energy (massless) limit of such amplitude. For the first and last term, only one of the transverse helicities of the $W$ boson are reached in this limit,
\al{\label{eq:dipole_massless_limit}
\frac{y_L}{M} \braket{\bm{13}} \braket{\bm{23}} & \to \frac{y_L}{M}\braket{13} \braket{23} = \mathcal{A}\left[1_\nu^- 2_{\bar{e}}^- 3_W^{-1}\right] ,\\
\frac{y_R}{M}\sbraket{\bm{13}} \sbraket{\bm{23}} & \to \frac{y_R}{M}\sbraket{13} \sbraket{23} = \mathcal{A}\left[1_\nu^+ 2_{\bar{e}}^+ 3_W^{+1}\right] .
}
Using the explicit formulas of App.~\ref{app:Dirac}, we see that these terms correspond, in the usual field theoretical language, to dipole operators involving the field strength $W^a_{\mu\nu}$. On the contrary, the high energy limit of the second and third term in Eq.~\eqref{eq:3-point_noflavor} is richer, since both the positive and negative helicities of the $W$ boson can be reached in this case. To be more precise, the correct form of the amplitude in the UV can be approached allowing the masses to vanish one at a time. Since the result will be important in Sec.~\ref{sec:PMNS}, it is worth to show it in detail. We have
\al{\label{eq:massless_limit}
\frac{g_L}{m_W} \braket{\bm{13}} & \sbraket{\bm{23}}  \stackrel{m_1 \to 0}{\longrightarrow}\frac{g_L}{m_W} \braket{1\bm{3}} \sbraket{\bm{23}} \\
&\stackrel{m_2 \to 0}{\longrightarrow} \frac{g_L}{m_W} \braket{1\bm{3}} \sbraket{2\bm{3}} =   - g_L \frac{\braket{1 \bm{3} }^2 }{\braket{12}} =  g_L \frac{\sbraket{2 \bm{3}}^2 }{\sbraket{12}}, 
}
where in the last step we have used momentum conservation and to get rid of the factor of $m_W$ we used the Weyl equations (see App. \ref{app:conventions_spinor_helicity}). This justifies our choice of writing the coefficient as $g_L/m_W$ since the presence of the vector mass in the denominator is necessary for a consistent UV-IR matching within the SM. Analogous reasoning can be applied to the $\sbraket{\bm{13}} \braket{\bm{23}}$ term. Again comparing with App.~\ref{app:Dirac}, we see that these terms are generated by the usual renormalizable charged current interaction.

For ease of the reader acquainted with the usual Lagrangian approach, we determine the $SU(2)_L\times U(1)_Y$ operators that generate the terms in Eq.~\eqref{eq:3-point_noflavor}. They can be found in Tab.~\ref{tab:UV_origin}. 
\begin{table*}[tb]
\begin{center}
\begin{tabular}{c|c|c}
amplitude & $U(1)_{EM}$ theory & $SU(2)_L \times U(1)_Y$ theory \\[0.4ex]
\hline
$\braket{\bm{13}}\braket{\bm{23}} $ & $\bar{e}_R \,\sigma^{\mu\nu} \nu_L W_{\mu\nu}^-$ ($d=5$) & $\bar{e}_R \sigma^{\mu\nu} H^\dag \tau^a L W^a_{\mu\nu}$ ($d=6$) \\[0.4ex]
\hline
\multirow{2}{*}{$\sbraket{\bm{13}} \sbraket{\bm{23}}$}  & \multirow{2}{*}{$\bar{e}_L \sigma^{\mu\nu} \nu_R W_{\mu\nu}^-$ ($d=5$)} & $(\bar{L} H) \sigma^{\mu\nu} (H^T \epsilon \tau^a L^c) W_{\mu\nu}^a$ ($d=7$, Majorana) \\[0.4ex]
& &  $\bar{L} \sigma^{\mu\nu} \tau^a N_R \epsilon H^* W_{\mu\nu}^a$ ($d=6$, Dirac) \\[0.4ex]
\hline
$\braket{\bm{13}}\sbraket{\bm{23}}$ & $\bar{e}_L \gamma^\mu \nu_L W_\mu^-$ ($d=4$) & $\bar{L} \gamma^\mu \tau^a L W_\mu^a$ ($d=4$) \\[0.4ex]
\hline
\multirow{2}{*}{$\sbraket{\bm{13}} \braket{\bm{23}} $} & \multirow{2}{*}{$ \bar{e}_R \gamma^{\mu} \nu_R W_{\mu}^-$ ($d=4$)} & $\bar{e}_R \gamma^\mu (H^\dag \epsilon L^c) H^\dag \epsilon D_\mu H^*$ ($d=7$,  Majorana) \\[0.4ex]
& & $\bar{e}_R \gamma^\mu N_R H^\dag \epsilon D_\mu H^*$ ($d=6$, Dirac)
\end{tabular}
\caption{\label{tab:UV_origin} Amplitude/operators dictionary. In the $SU(2)_L \times U(1)_Y$ invariant theory we list the smallest dimensional operators contributing to the corresponding amplitude, distinguishing between the Majorana and Dirac cases, if needed. }
\end{center}
\end{table*}
For Majorana neutrinos we have $\nu_R = \nu_L^c$ (we do not add any additional sterile neutrino to the physical spectrum, {\it i.e.} these states lie above the cutoff of the EFT), while for Dirac neutrinos this is an independent degree of freedom $\nu_R = N_R$.  The table emphasises one of the crucial points raised in the Introduction: on-shell amplitudes automatically include structures that (i) appear at different order in an expansion over the cutoff scale and (ii) that traditionally belong to different EFTs: the SMEFT for the terms involving $L^c$ and the $\nu$SMEFT for the terms involving $N_R$~\footnote{Typically, in the $\nu$SMEFT the right handed neutrinos are supposed to be heavier than the left handed ones and are responsible for the generation of neutrino masses. Since we are considering Dirac neutrinos, we are implicitly assuming that the Majorana term for the $N_R$ fields is forbidden by some symmetry.}. Moreover, we observe from Tab.~\ref{tab:UV_origin} that the structures involving the right handed neutrino helicity are more suppressed for Majorana than for Dirac neutrinos when embedded in an EFT framework.
Can the same conclusion be reached from a purely on-shell perspective? An apparent obstacle to this program is the fact that only the $\braket{\bm{13}} \sbraket{\bm{23}}$ structure can be directly UV completed into the $SU(2)_L \times U(1)_Y$ invariant 3-point amplitudes (see Eq.~\eqref{eq:SM_amplitudes}). According to Tab.~\ref{tab:UV_origin} all other terms should be generated by higher-point amplitudes. Two arguments can be used to perform the IR/UV matching between amplitudes involving a different number of particles. The first one was presented in Ref.~\cite{Durieux:2019eor}, and amounts to notice that, in the soft limit in which the Higgs boson momentum vanishes, amplitudes involving the Higgs boson are indistinguishable from amplitudes without the Higgs boson. To compensate for the dimension mismatch between higher and lower points amplitudes, a new mass scale must be introduced, analogous to the Higgs vev. A second argument has been presented in Ref.~\cite{Bachu:2019ehv}  and amounts to impose correlations between coefficients to tame a possible growth with energy of the amplitude. In our case this procedure can be applied to the $\sbraket{\bm{13}} \braket{\bm{23}}$ structure. Focussing first on Dirac neutrinos, we need to ``glue'' the 3-point amplitudes $\nu\bar{e}W^\pm$ and $h W^\pm W^0$ (with $W^0$ the longitudinal component of the $W$ boson and $W^\pm$ the transverse ones) to obtain the 4-point amplitude $\nu \bar{e} W^0 H$. This amplitude can be UV completed in the $SU(2)_L \times U(1)_Y$ invariant amplitude $\nu \bar{e} H^\dag H$, whose coefficient is given at leading order by $1/\Lambda^2$. Demanding this 4-point amplitude not to grow with the energy requires $g_R \sim m_W^2/\Lambda^2$, apart from an $\mathcal{O}(1)$ coefficient. A similar reasoning applies to the Majorana case, in which however we need to construct the 5-point amplitude $L^c \bar{e} H^* H^\dag H$ to obtain an $SU(2)_L \times U(1)_Y$ invariant object. This generates a dependence $g_R \sim \Lambda^{-3}$. There is, nevertheless, an obstruction to applying the same procedure to the dipole amplitudes, since these are generated at loop level. Since this point lies somewhat outside the purpose of this paper, we defer it to future work. In what follows, we will never need the cutoff scaling of the terms in Eq.~\eqref{eq:3-point_noflavor}. 

\subsection{\label{sec:Emergence} Emergence of the PMNS matrix}

Up to this point, we have not considered how flavor can be implemented in the 3-point amplitude. Clearly, we could invoke the operators in Table~\ref{tab:UV_origin} and extract the flavor structure of the amplitudes directly from them, but in this work we want to avoid the use of QFT techniques.
Our derivation rests on one important assumption: in the absence of mass, no quantum number can be used to distinguish between 1-particle states of different generations. In the massless limit we thus gain the freedom to perform unitary transformations on the states of each species, which amount to
\al{\label{eq:flavor_tr}
\ket{\nu_i(\bm{p}, h)} & \to (U_\nu^*)_{ji} \ket{\nu_j(\bm{p}, h)}, \\
\ket{\bar{e}_i( \bm{p}, h)} & \to (U_e)_{ji} \ket{\bar{e}_j( \bm{p}, h)},
}
where $U_\nu$ and $U_e$ are unitary matrices~\footnote{We take the usual transformation in which antiparticles transform in the conjugate representation with respect to particles.}. The transformations above imply that in the massless limit the corresponding $S$-matrix elements must transform covariantly under flavor transformations. At the level of amplitudes, massless spinor variables depend only on the particle momentum and are thus generation blind. As a consequence, all the flavor dependence must be encoded in the coefficients in front of each term in the amplitude, which must thus have non-trivial flavor transformations. Stated in another way: when a certain type of particle becomes massless, the amplitude must be a covariant tensor under a flavor transformation of that type of particle. Notice that this is not the case for massive amplitudes, in which also the spinor variables depend on the particle mass.

We now use this observation to deduce the flavor structures of the coefficients $g_{L,R}$ and $y_{L,R}$. More specifically, we take the limit in which individual particles become massless one at a time: first all neutrinos, then the charged leptons and finally the $W$ boson. When all the particles are massless we will match into the SM gauge amplitude~\cite{Christensen:2018zcq,Baratella:2020lzz}:
\al{\label{eq:SM_amplitudes}
\mathcal{A}_{SM}\left[1_{L_{A,i}}^- 2_{\bar{L}_{B,j}}^+ 3_{W^a}^{-1} \right] & = g^{ij} (T^a)_{AB} \frac{\braket{13}^2}{\braket{12}}, \\
\mathcal{A}_{SM}\left[1_{L_{A,i}}^- 2_{\bar{L}_{B,j}}^+ 3_{W^a}^{+1} \right] & = - g^{ij} (T^a)_{AB} \frac{\sbraket{23}^2}{\sbraket{12}},
}
where $L$ is the usual lepton doublet and $T^a$ is a gauge generator. We have also written explicitly the gauge indices ($A$ and $B$) as well as the generation indices ($i$ and $j$). Under the assumption of flavor universality, we can use the freedom to rotate $L_{A,i}$ and $L_{B,j}$ to make the $g^{ij}$ coefficients proportional to the identity: $g^{ij} \to g \delta^{ij}$, where $g$ is  the $SU(2)_L$ gauge coupling. 

We are now in the position of discussing the flavor dependence of the couplings in the IR. To fix our notation, we rewrite Eq.~\eqref{eq:3-point_noflavor} making explicit the generation indices:
\al{\label{eq:3-point_flavor}
\mathcal{A}\left[ 1_{\nu_i} 2_{\bar{e}_j} 3_W \right] & = \frac{y_L^{ij}}{M} \braket{\bm{1}_i\bm{3}} \braket{\bm{2}_j\bm{3}} + \frac{g_L^{ij}}{m_W} \braket{\bm{1}_i\bm{3}} \sbraket{\bm{2}_j\bm{3}} \\
& + \frac{g_R^{ij}}{m_W} \sbraket{\bm{1}_i\bm{3}} \braket{\bm{2}_j\bm{3}} + \frac{y_R^{ij}}{M} \sbraket{\bm{1}_i\bm{3}} \sbraket{\bm{2}_j\bm{3}} .
}
Since we are considering on-shell amplitudes, the generation indices represent states of well-defined mass, and all the coefficients are complex matrices in generation space. Before taking the massless limit, we observe that, according to Eqs. \eqref{eq:dipole_massless_limit}, \eqref{eq:massless_limit} and \eqref{eq:SM_amplitudes}, only the $\braket{\bm{1}_i\bm{3}} \sbraket{\bm{2}_j\bm{3}}$ term has the correct particle content to be matched into the SM amplitudes. We thus need to show that the matrix $g_L$ is not any complex matrix, but is unitary and can thus be identified with the PMNS matrix. The various massless limits can be achieved following Eq.~\eqref{eq:massless_limit}. At each stage we gain the freedom to perform a flavor transformation of the species that became massless. To book keep this freedom, we explicitly apply a generic flavor transformation at each stage. Focusing on the coefficients only and writing them as matrices in generation space we obtain
\be\label{eq:flavor_matching}
\frac{g_L}{m_W}  \stackrel{m_\nu \to 0}{\longrightarrow} \frac{U_\nu^\dag g_L}{m_W}  \stackrel{m_e \to 0}{\longrightarrow} \frac{U_\nu^\dag g_L U_e}{m_W}  \stackrel{m_W \to 0}{\longrightarrow} U_\nu^\dag g_L U_e = g \bm{1} ,
\ee
where in the last step we have matched onto the SM amplitude with coefficient proportional to the identity in flavor space. The last equality can be true only if $g_L$ is proportional to a unitary matrix. We will thus make contact with the usual notation and write
\be\label{eq:PMNS}
g_L = g \,U_{\nu} U_{e}^{\dag} \equiv g U_{\rm{PMNS}}.
\ee
The PMNS matrix emerges naturally from our considerations. As $g_L$ is proportional to a unitary matrix, we observe that as soon as the neutrinos  become massless, we can use the freedom of rotating the neutrino states to make $g_L$ proportional to the identity. This matches, as it should, the usual QFT conclusion that no PMNS matrix appears in the limit of massless neutrinos. Furthermore, it is interesting to observe that the standard counting of PMNS parameters is guaranteed thanks to the possibility of applying arbitrary phase transformations to the neutrino and charged antilepton 1-particle states. For Dirac neutrinos the pure phase transformations can be deduced directly from Eq.~\eqref{eq:flavor_tr} by identifying $(U_\nu)_{ij} \equiv e^{i \alpha_i} \delta_{ij}$ and $(U_e)_{ij} \equiv e^{i\beta_i} \delta_{ij}$. After this step the counting of phases can proceed as usual \cite{Giganti_2018}. The situation is different for Majorana neutrinos, since in this case particle and antiparticle coincide. Given that the latter transforms in the conjugate representation, consistency is ensured requiring $U_\nu = U_\nu^*$, i.e. the transformation is constrained to be orthogonal. This means that no phase transformation can be applied in a consistent way on the Majorana neutrino 1-particle states, leaving us with 3 phases in $U$.  Once more, we recover the usual parameter counting for the PMNS matrix. 

When the other terms in Eq.~\eqref{eq:3-point_flavor} are turned on, a similar reasoning applies, with a crucial difference: since they cannot be matched into any SM 3-point amplitude, we cannot conclude that the $y_{L,R}$ and $g_R$ coefficients are unitary. They will thus be generic $3\times 3$ complex matrices in flavor space, with $9 \times 2 = 18$ parameters each. If we choose to work in a basis in which the PMNS matrix $g_L$ has the minimum number of parameters, no freedom is left to reduce the number of parameters of the $y_{L,R}$ and $g_R$ matrices, in such a way that the amplitude of Eq.~\eqref{eq:3-point_flavor} depends on $18 \times 3 = 54$ parameters in addition to those appearing in the PMNS matrix. Only the number of parameters of the latter depends on the Dirac or Majorana nature of the neutrinos, while in both cases the remaining coefficients will depend on a total of 54 parameters.

Having established how the PMNS matrix emerges, the usual oscillation formula in vacuum can now be obtained considering two amplitudes ${\cal A}_P$ and ${\cal A}_D$, each containing one neutrino state and denoting, respectively, the production and detection process, connected by a neutrino propagator. The overall amplitude can be computed in the limit of on-shell propagating neutrinos using the usual factorisation properties~\cite{Arkani-Hamed:2017jhn}. We will not show explicitly this computation here, since the exact form of the expressions depends on the chosen production/detection processes.

We conclude this section with two observations: first, the amplitude for a $Z$ boson interacting with neutrinos can be directly obtained from equation \eqref{eq:flavor_matching} with the replacement $e \to \nu$. This implies that $g_L = g U_\nu U_\nu^\dagger = g\bm{1}$, hence, as expected, this term is flavor blind. Moreover, a line of reasoning  similar to the one used above can be applied to charged current interactions between quarks. Although the quark and $W$ mass hierarchy forbids separate massless limits, the UV matching can be done taking all the particles massless at once. We again conclude that the coefficient of the $\braket{\bm{1}_i\bm{3}} \sbraket{\bm{2}_j\bm{3}}$ term must be proportional to a unitary matrix, to be identified with the CKM matrix. The main difference with respect to the present case is the UV origin of the other three terms \footnote{For completeness we report here the operators that generate the various terms in the $\mathcal{A}\left[1_u 2_{\bar{d}} 3_W\right]$ amplitude: the term $\braket{\bm{13}} \braket{\bm{23}}$ is generated at leading order by the operator $\bar{d}_R \sigma^{\mu\nu} \tau^a H^\dag Q W_{\mu\nu}^a$; $\sbraket{\bm{13}} \sbraket{\bm{23}}$ by $\bar{Q} \tilde{H} \sigma^{\mu\nu} \tau^a u_R W_{\mu\nu}^a$; $\braket{\bm{13}} \sbraket{\bm{23}}$ by the SM operator $\bar{Q} \gamma^\mu \tau^a Q W_\mu^a$; finally, $\sbraket{\bm{13}} \braket{\bm{23}}$ is generated by $\bar{d}_R \gamma^\mu u_R H^\dag \epsilon D_\mu H^*$.}.

\vspace{1em}

\section{\label{sec:oscillations} Oscillations in matter}


Having established the flavor structure of the scattering amplitudes we are ready to tackle the problem of neutrino oscillations in matter. As already mentioned, in the usual approach matter effects are computed using Hamiltonians, a tool we do not have at our disposal when studying the problem from a scattering amplitude point of view. How can we thus describe such phenomenon?

\subsection{\label{sec:potential} The potential}

Firstly, we have to translate the usual computation of the matter potential in the scattering amplitude language. To this end, we consider the elastic scattering $\nu e \to \nu e$ mediated by a charged current, focusing for the moment on the SM only contribution. This can be constructed using the usual factorisation properties around the pole. Using the SM term in the amplitude of Eq.~\eqref{eq:3-point_flavor} we find
\begin{widetext}
\al{\label{eq:factorization}
\adjustbox{valign=m}{
 \begin{tikzpicture}[line width=0.75] 
\draw[f] (-0.75, 0.75) node[left]{$4_{\nu_{i'}}$} -- (0,0) node[midway, above]{$\searrow$};
\draw[f]  (0,0)   -- (-0.75, -0.75) node[left]{$3_{\bar{e}_{j'}}$} node[midway,below]{$\nearrow$};
\draw[v] (0.75,0) -- (0,0) node[midway,above]{$\stackrel{P}{\rightarrow}$} node[midway,below]{$W$};
\draw[f] (0.75,0) -- (1.5, 0.75) node[right]{$2_{\bar{\nu}_i}$} node[midway,above]{$\swarrow$};
\draw[f] (1.5, -0.75) node[right]{$1_{e_j}$} -- (0.75, 0) node[midway,below]{$\nwarrow$};
\end{tikzpicture} }={\cal A}\left[1_{e_j} 2_{\bar{\nu}_i} 3_{\bar{e}_{j'}} 4_{\nu_{i'}}\right] =  -\frac{g^2 U^{ij} (U^{i' j'})^*}{P^2 - m_W^2}\left[ - \braket{ {\bf 2}_i {\bf 3}_{j'} }  \sbraket{{\bf 1}_j {\bf 4}_{i'} } + \frac{1}{m_W^2} \asbraket{{\bf 2}_i}{P}{{\bf 4}_{i'}} \asbraket{{\bf 3}_{j'}}{P}{{\bf 1}_j}
\right] ,
}
\end{widetext}
with $U\equiv U_\text{PMSN}$. As discussed before, we will not consider neutral currents, since the contribution is not generation dependent. This would lead to the same shift in the neutrino mass regardless of the neutrino flavor. We now focus on the low energy limit of the amplitude ($P \to 0$) in order to recover the usual MSW potential~\cite{Wolfenstein:1977ue,Mikheyev:1985zog}. We first take the elastic limit $p_3 = - p_1$ and $p_2 = - p_4$ and impose $j=j'$ \footnote{To obtain the elastic amplitude we would in principle need to take $i=i'$ as well. However, since $|m_{\nu_i}-m_{\nu_{i'}}|\ll m_{e_j}$, there is not enough energy resolution to distinguish between neutrino mass states, hence they contribute coherently to the amplitude.}:
\be\label{eq:amplitude_before_spin_average}
{\cal A}\left[1_{e_j} 2_{\bar{\nu}_i} 3_{\bar{e}_{j'}} 4_{\nu_{i'}}\right] \to - \frac{g^2 U^{ij} (U^{i' j})^*}{m_W^2} \braket{ {\bf 4}_i {\bf 1}_{j} }  \sbraket{{\bf 1}_j {\bf 4}_{i'} }.
\ee
We then perform a spin average over the electron spin. To this end, we remind the reader that the spinor helicity part of the previous amplitude is written, without the bold notation, as $\braket{4_i^J 1_{j'}^K} \sbraket{1_j^I 4_{i'}^L}$, where $\left\{I, J, K, L \right\}$ are $SU(2)$ little group indices. Taking the average over the electron spin thus amounts to contract the amplitude in Eq.~\eqref{eq:amplitude_before_spin_average} by $\epsilon_{IK}/2$. This leads to
\be\label{eq:amp_4_particle}
\overline{{\cal A}}\left[1_{e_j} (-4)_{\bar{\nu}_i} (-1)_{\bar{e}_{j}} 4_{\nu_{i'}}\right] = 2 \sqrt{2} G_F U^{ij} (U^*)^{i'j} \asbraket{4_i^J}{p_{e_j}}{4_{i'}^L}
\ee
where $\overline{\cal A}$ denotes the averaged amplitude and $G_F = \sqrt{2} g^2/8 m_W^2$. We have also explicitly put in evidence the charged lepton momentum $p_{e_j} = \ket{1_j^I} \sbra{1_{jI}}$. The MSW potential $V_\text{MSW}^{(j)}$ with respect to a medium containing charged leptons $e_j$ is defined, in the usual field theoretical derivation, as ${\cal L}_{int} = \bar{\nu} \gamma^\mu \nu V_{\text{MSW},\mu}$. Interpreting this in the amplitude language, we define $V_{\rm MSW}^{(j)}$ as the amplitude~\eqref{eq:amp_4_particle} stripped of the neutrino spinor helicity variables and integrated over the charged lepton Lorentz-invariant momentum phase space:
\be \label{eq:VMSW}
(V_{\rm MSW}^{(j)})^{ii'}= \sqrt{2} G_F U^{ij} (U^*)^{i'j} N_{e_j} \left\langle \frac{p_{e_j}}{E_{e_j}} \right\rangle ,
\ee
where we have defined the average over the medium
\be
\left\langle A \right\rangle = \frac{1}{N_{e_j}} \int \frac{d^3p_{e_j}}{(2\pi)^3} f(\bm{p}_{e_j}) A
\ee
using the medium distribution function $f(\bm{p}_{e_j})$ and the number of medium constituent $N_{e_j}$ is given by
\be
N_{e_j} = \int \frac{d^3p_{e_j}}{(2\pi)^3} f(\bm{p}_{e_j}) .
\ee
This expression agrees with the results in the literature, see for instance Ref.~\cite{Bergmann:1999rz}. We also observe that the potential extracted from the amplitude \eqref{eq:amp_4_particle} is the one felt by neutrinos over a charged lepton background. To obtain the same potential for anti-neutrinos, it is enough to change the direction of the neutrino momentum, which implies in a sign change due to $\sket{\bm{4}_{i'}} \rightarrow \sket{-\bm{2}_{i'}} = - \sket{\bm{2}_{i'}}$ (see App. A). Hence, the analytic continuation conditions of the spinor variables guarantee that the MSW potential for neutrinos and anti-neutrinos have opposite signs.

As a further check, let us verify what happens considering the exchange of a spin-0 particle (called $\phi$) instead of a spin-1 particle. The 3-point amplitude is in this case
\be
{\cal A}[1_{\bar{f}} 2_f 3_\phi] = c_1^f \braket{\bf 12} + c_2^f \sbraket{\bf 12},
\ee
where $f =\nu, e$. This leads to a 4-point amplitude ${\cal A}[1_e 2_{\bar{\nu}} 3_{\bar{e}} 4_\nu]$ which, once more, needs to be computed in the elastic limit and must be mediated over the electron spins. A crucial difference with respect to the previous case is that, while in the case of spin-1 mediator, the spin average resulted in a factor of the medium particle momentum, $p_{e_j} = \ket{1_j^I} \sbra{1_{jI}}$, in the scalar case we can use the identity $\epsilon_{IJ} \braket{1^I 1^J} = m_e$, introducing terms proportional to the medium particle mass. The final result is of the form 
\be\label{eq:scalar_mediator}
\overline{\cal A}[1_e 2_{\bar{\nu}} (-1)_{\bar{e}} (-2)_\nu] = m_e \left( C_1 \braket{\bf 2 2} - C_2 \sbraket{\bf 22}\right),
\ee
where $C_{1,2}$ are appropriate combinations of the $c_{1,2}^{\nu, e}$ coefficients. The matter potential in this case is thus proportional to the medium particle mass $m_e$, once more the same result found in the literature~\cite{Bergmann:1999rz}.

\vspace{1em}
\subsection{\label{sec:propagation} Effect on neutrino propagation}

In possession of the matter potential, we are now in position to propose a strategy to recover a modified dispersion relation using on-shell amplitudes. We will devote this section to the SM case, leaving to Sec.~\ref{sec:BSM} the discussion of BSM effects. In the standard approach, this step is accomplished including the matter potential in the Hamiltonian (or in the equation of motion), followed by a rotation to go to the matter basis, which defines the in-medium propagating eigenstates. Using only amplitudes, we instead propose to consider a pair of generic $n$-point amplitudes ${\cal A}_P[ 1_{\bar{\nu}} \left\{ {\cal I} \right\}]$, ${\cal A}_D[1_\nu \left\{ {\cal I}' \right\} ]$, containing one outgoing and one incoming neutrino, respectively. The collective indices $\left\{ {\cal I} \right\}$ and $\left\{ {\cal I}' \right\}$ represent some collections of particles other than the neutrinos. The amplitudes ${\cal A}_P$ and ${\cal A}_D$ will be used to represent neutrino production and detection, respectively. For instance, they may represent 4-particle amplitudes containing a charged lepton and a nucleon pair, but we will not need to specify their form in our discussion. For ease of notation, we will from now on use the simplified notation ${\cal A}_P[ 1_{\bar{\nu}} \left\{ {\cal I} \right\} ] = {\cal A}_P[\bar{\nu}_i^I]$ and ${\cal A}_D[1_\nu \left\{ {\cal I}' \right\} ] = {\cal A}_D[\nu_j^J]$, leaving implicit the information about the additional particles that will not play a role in our discussion and stressing the relevant indices of the neutrino state: its massive little group indices $I$ and $J$ and its mass flavor indices $i$ and $j$. 
Using these amplitudes we can construct a 4-point amplitude with a neutrino factorisation channel, which would account for its propagation in vacuum from the production to the detection points. In the presence of a charged lepton background, we can build higher-point functions by allowing the internal neutrino to interact with a pair of external charged leptons. Considering that we are in the elastic limit we can write the series

\begin{widetext}
\al{\label{eq:series_potential}
{\cal A}_{\rm prop}\left[P\rightarrow D\right] & = 
\adjustbox{valign=m}{
 \begin{tikzpicture}[line width=0.75] 
\draw (-1.5,0.75) -- (-0.75, 0);
\draw (-1.5, 0) node[left]{${\cal A}_P$}-- (-0.75, 0);
\draw (-1.5, -0.75) -- (-0.75, 0);
\draw[f] (-0.75, 0) -- (0, 0) node[midway, above]{$\nu$};
\draw (0,0) -- (0.75, 0.75);
\draw (0,0) -- (0.75,0) node[right]{${\cal A}_D$};
\draw (0,0) -- (0.75, -0.75);
\end{tikzpicture} } + 
\adjustbox{valign=m}{
 \begin{tikzpicture}[line width=0.75] 
\draw (-1.5,0.75) -- (-0.75, 0);
\draw (-1.5, 0) node[left]{${\cal A}_P$}-- (-0.75, 0);
\draw (-1.5, -0.75) -- (-0.75, 0);
\draw[f] (-0.75, 0) -- (0, 0) node[midway, above]{$\nu$};
\draw[f] (0,0) -- (0.75,0) node[midway, above]{$\nu$};
\draw (0+0.75,0) -- (0.75+0.75, 0.75);
\draw (0+0.75,0) -- (0.75+0.75,0) node[right]{${\cal A}_D$};
\draw (0+0.75,0) -- (0.75+0.75, -0.75);
\filldraw[black] (0,0) circle (2pt);
\filldraw[black] (0, -0.75) circle (2pt);
\draw[dashed] (0,0) -- (0, -0.75);
\end{tikzpicture} } + \dots \\
& = \sum_{i,j} {\cal A}_P[\bar{\nu}_i^I]\frac{1}{p^2 - m_{\nu_i^2}}  \left[\delta^{ij} \epsilon_{IJ}+ \frac{ \asbraket{p_{iI}}{V_{\rm MSW}}{p_{jJ}}}{p^2 - m_{\nu_j}^2} + \dots \right] {\cal A}_D[\nu_j^J] ,
}
\end{widetext}
where the line \begin{tikzpicture} \filldraw[black] (0,0) circle (2pt); \draw[dashed] (0,0) -- (0.75,0); \filldraw[black] (0.75,0) circle (2pt); \end{tikzpicture} represents an insertion of the amplitude in Eq.~\eqref{eq:amp_4_particle} containing the MSW potential. We mimic in this way the fact that neutrinos have interactions as they propagate through matter. To compare with the literature, we consider only the  contributions of order $\mathcal{O}(G_F)$ to the series. In order to proceed, it is important to understand the behaviour of the $\asbraket{p_{iI}}{V_{\rm MSW}}{p_{jJ}}$ insertion. We will consider ultra relativistic neutrinos and investigate the high energy limit. The discussion is more transparent by using the high energy basis described in~\citep{Arkani-Hamed:2017jhn, Bachu:2019ehv} (see App. \ref{app:conventions_spinor_helicity} for details), such that  
\begin{align}\label{eq:hel}
	\begin{split}
		&\bra{p_{i,I}}(V_{\rm{MSW}})^{ij}\sket{p_{j,J}} =\\
		&\bra{\lambda_{i}}(V_{\rm{MSW}})^{ij}|\tilde{\lambda}_{j}]		
		\xi^{-}_I\xi^{+}_J + \bra{\lambda_{i}}(V_{\rm{MSW}})^{ij}	  		\sket{\tilde{\eta}_{j}}\xi^{-}_I\xi^{-}_J\\
		&+ \bra{\eta_{i}}(V_{\rm{MSW}})^{ij}|\tilde{\lambda}_{j}]	
		\xi^{+}_I\xi^{+}_J
		+
		\bra{\eta_{i}}(V_{\rm{MSW}})^{ij}\sket{\tilde{\eta}_{j}}		
		\xi^{+}_I	
		\xi^{-}_J.
	\end{split}
\end{align}
In the limit $m_{\nu} \to 0$ only the first term contributes, selecting the massive little group indices $I=2$ and $J=1$. This is compatible with the structure of the production and detection amplitudes, since ${\cal A}_P[\bar{\nu}_i^I]$ must contain an ingoing antineutrino state, {\cal i.e.} a spinor helicity variable $\sket{p_i^I}$ that, in the high energy limit, selects $I=2$, while the ${\cal A}_D[\nu_j^J]$ contains an incoming neutrino state, {\it i.e.} a spinor helicity variable $\ket{p_j^J}$ that in the massless limit, selects $J=1$. Physically, this means that the propagating neutrino has negative helicity, {\it i.e.} it has left handed chirality as expected in the SM case we are considering. 
In light of this, we only need to analyse $\asbraket{p_{i2}}{V_{\rm MSW}}{p_{j1}}$ \footnote{We have checked that all other combinations of little group indices are suppressed by at least another power of the neutrinos mass, making them even more suppressed.}. Using the explicit expressions for the massive spinor variables presented in App. \ref{app:conventions_spinor_helicity} we obtain
\be
\sket{p_{j1}}^{\dot{\alpha}}\bra{p_{i2}}^{\alpha} =  p^{\dot{\alpha}\alpha} + \order{m_\nu^2/|\bm{p}|^2} ,
\ee
from which, neglecting terms of order $\mathcal{O}(m_{\nu}^{2})$, we finally get
\be
\asbraket{p_{i2}}{V_{\rm MSW}}{p_{j1}} =  2 p_\mu V_{\rm MSW}^\mu,
\ee
with the whole generation structure encoded in the potential. 

Going back to the propagation amplitude, we can thus write, using a matrix notation,
\begin{widetext}
\al{\label{eq:series_resum}
{\cal A}_{\rm prop}\left[P\rightarrow D\right] & =   {\cal A}_P[\bar{\nu}^2]\frac{1}{p^2 - {\cal M}_{\nu}^2}  \left[\bm{1} + 2p\cdot V_{\rm MSW} \frac{1}{p^2 - {\cal M}_{\nu}^2} + \dots \right]{\cal A}_D[\nu^1]\\
& =  {\cal A}_P[\bar{\nu}^2]\frac{1}{\left(p - V_{\rm MSW}\right)^2 - {\cal M}_{\nu}^2}{\cal A}_D[\nu^1],
}
\end{widetext}
up to terms of order $\order{G_F^2}$. In the last line of the equation above we obtain the modified dispersion relation for the neutrinos, \textit{i.e.} a shift to the 4-momentum which agrees with the literature \cite{Bergmann:1999rz}.

 The re-summed amplitude is not in an explicit factorisable form, as the neutrino states mix due to the flavor off-diagonal terms in the potential. In order to restore manifest factorisation, \textit{i.e.} to respect locality, we must be able to write Eq. \eqref{eq:series_resum} as a single sum over well defined channels. This amounts to require that the matrix $\left(p-V_{\rm MSW}\right)^2-{\cal M}_\nu^2$ must be diagonalizable. Calling $U_M$ the unitary matrix that diagonalizes the propagator in matter, we obtain
\be\label{eq:matter_basis}
{\cal A}_{\rm prop}\left[P\rightarrow D\right] =  \tilde{\cal A}_P[\bar{\nu}^2]\frac{1}{p^2 - \tilde{\cal M}_{\nu}^2}\tilde{\cal A}_D[\nu^1],
\ee
with $\tilde{\cal A}_P[\bar{\nu}^2] \equiv {\cal A}_P[\bar{\nu}^2] U_M^\dagger$, $\tilde{\cal A}_D[\nu^1] \equiv U_M {\cal A}_D[\nu^1]$ and $\tilde{\cal M}_{\nu}$ the new diagonal mass matrix. The transition from Eq. \eqref{eq:series_resum} to \eqref{eq:matter_basis} is nothing but the diagonalization of the vacuum mass eigenbasis to the matter one.

To conclude this section, and as a sanity check of the validity of our procedure, we perform the same computation for the potential generated by a scalar mediator. We take from Eq. \eqref{eq:scalar_mediator} the potentials associated to the terms with angle and square brackets,
\be
V_{\phi,1} = m_e C_1,\quad V_{\phi,2} = m_e C_2 ,
\ee
and we repeat the same procedure outlined before, noticing that now $\braket{p_{i2}p_{j1}}=m_{\nu_i}$, $\sbraket{p_{i2}p_{j1}}=-m_{\nu_j}$ to first order in $m_\nu^2/\bm{p}^2$. The modified dispersion relation reads
\be\label{eq:scalar_dispersion}
p^2-{\cal M}_\nu^2\rightarrow p^2-{\cal M}_\nu^2 - {\cal M}_\nu V_{\phi,1}-V_{\phi,2}{\cal M}_\nu ,
\ee
{\it i.e.} the scalar mediator induces a change in the neutrino mass matrix that is proportional to the mass of the background constituents. We again recover the results in the literature.

\section{\label{sec:BSM} BSM effects}

We now turn to the computation of the matter effects using the most general 4-point function ${\cal A}\left[1_{e_j} 2_{\bar{\nu}_i} 3_{\bar{e}_{j}} 4_{\nu_{i'}}\right]$ that can include any possible BSM physics. Here we will confine ourselves to a scenario in which no sterile neutrino is produced by ${\cal A}_P$ and consider only the propagation of the SM neutrinos. This means that the amplitude ${\cal A}\left[1_{e_j} 2_{\bar{\nu}_i} 3_{\bar{e}_{j}} 4_{\nu_{i'}}\right]$ will represent the most general correction to the SM propagation. According to Ref. \cite{Durieux:2020gip}, this amplitude can be written as
\begin{widetext}
\al{\label{eq:BSM_amp}
{\cal A}\left[1_{e_j} 2_{\bar{\nu}_i} 3_{\bar{e}_{j'}} 4_{\nu_{i'}}\right] & =  g_1\braket{\bm{2}_{i}\bm{4}_{i'}}\braket{\bm{1}_{j}\bm{3}_{j'}}+g_2\braket{\bm{2}_{i}\bm{4}_{i'}}\sbraket{\bm{1}_{j}\bm{3}_{j'}}+g_3\braket{\bm{1}_{j}\bm{2}_{i}}\sbraket{\bm{3}_{j'}\bm{4}_{i'}}+g_4\braket{\bm{2}_{i}\bm{3}_{j'}}\sbraket{\bm{1}_{j}\bm{4}_{i'}}+g_5\sbraket{\bm{1}_{j}\bm{3}_{j'}}\asbraket{\bm{2}_{i}}{\bm{1}}{\bm{4}_{i'}}+\\
& + g_6\sbraket{\bm{1}_{j}\bm{3}_{j'}}\sabraket{\bm{2}_{i}}{\bm{1}}{\bm{4}_{i'}}+g_7\braket{\bm{2}_{i}\bm{4}_{i'}}\sabraket{\bm{1}_{j}}{\bm{2}}{\bm{3}_{j'}}+g_8\braket{\bm{2}_{i}\bm{4}_{i'}}\asbraket{\bm{1}_{j}}{\bm{2}}{\bm{3}_{j'}} + ({\rm angle}\leftrightarrow{\rm square}),
}
\end{widetext}
where the structures with the conjugate spinors $({\rm angle}\leftrightarrow{\rm square})$ may have independent coefficients. Also, all coefficients may depend on the Mandelstam variables $s_{ij}=(p_i+p_j)^2$. The amplitude above spans the most general 4-point amplitude, which includes both factorisable and contact terms. We also note that, except for the SM term $g_4$ that is proportional to $U^{ij}(U^{i'j'})^*$, all other coefficients are arbitrary matrices in generation space.
Taking the elastic limit and averaging over the charged lepton background yields the amplitude
\al{\label{eq:BSM_reduced_amp}
{\cal A}\left[1_{e_j} 2_{\bar{\nu}_i} 3_{\bar{e}_{j'}} 4_{\nu_{i'}}\right] & \rightarrow V_1 \braket{4^J_{i}4^L_{i'}} - V_2 \sbraket{4^J_{i}4^L_{i'}} + \\
& + \asbraket{4^J_{i}}{V_3}{4^L_{i'}},
}
where the different potentials read
\be\label{eq:V1}
V_1  =N_{e_j}\left\langle \frac{(p_4\cdot p_{e_j})(g_7+g_8)-m_{e_j}(g_1+g_2)}{2E_{e_j}}\right\rangle,
\ee
\be\label{eq:V2}
V_2  =N_{e_j}\left\langle \frac{(p_4\cdot p_{e_j})(g_7'+g_8')-m_{e_j}(g_1'+g_2')}{2E_{e_j}}\right\rangle,
\ee
\be\label{eq:V3}
V_3  =-\frac{N_{e_j}}{2}\left\langle \left[(g_4-g_3)+2m_{e_j}(g_6-g_5)\right]\frac{p_{e_j}}{2E_{e_j}}\right\rangle.
\ee
The symbols $g'$ denote the coefficients for the conjugate spinor structures. We see that in Eq. \eqref{eq:BSM_reduced_amp} there is no contribution such as $\sabraket{4^J_{i}}{V}{4^L_{i'}}$ because it is  higher-order in the neutrino mass expansion. Using the results from Eqs. \eqref{eq:series_resum} and \eqref{eq:scalar_dispersion}, we obtain the most general modification to the neutrinos dispersion relations up to first order in the coefficients:
\be\label{eq:BSM_dispersion}
p^2-{\cal M}_\nu^2\rightarrow \left(p-V_3\right)^2-{\cal M}_\nu^2 - {\cal M}_\nu V_{1}-V_{2}{\cal M}_\nu.
\ee
In order to have a well defined factorisable amplitude, \textit{i.e.} to define a matter basis for the neutrino propagation, Eq. \eqref{eq:BSM_dispersion} must be a diagonalizable matrix. 


From a broader phenomenological point of view, Eq.~\eqref{eq:BSM_amp} is the on-shell version of amplitudes generated by neutrino Non-Standard Interactions (NSI), including interactions of right handed neutrinos~\cite{Proceedings:2019qno}. The amplitude's coefficients can thus in principle be probed in a variety of experiments. For instance, the process $\nu e^- \to \nu e^-$ has been searched for at neutrino scattering experiments, while $e^+e^- \to \nu \bar{\nu}$ can be studied at colliders (like LEP, Babar and Belle) and can also give an additional contribution to {supernov\ae} energy draining~\cite{Workman:2022,Stapleford:2016jgz}. In any case, there are not enough data to probe all the parameters appearing in Eq.~\eqref{eq:BSM_amp}. Another interesting feature of the amplitude \eqref{eq:BSM_amp} lies in the difference between Dirac and Majorana neutrinos. In the latter case, assuming all coefficients to be constant and taking the incoming and outgoing neutrino indices to be the same, the amplitude must be invariant under a $2 \leftrightarrow 4$ exchange. This reflects in the following correlations between the coefficients: $g_{3,4} = -g_{4,3}'$, $g_{5,6}^{(\prime)} = -g _{6,5}^{(\prime)}$ and
\al{\label{eq:coeffs_Majorana}
g_1 = &  -g_1 - g_7 m_j + g_8 m_{j'}, \\
g_2 = & - g_2+ g_7 m_{j'} - g_8 m_j,  \\
g_1' = & - g_1' + g_7' m_j - g_8' m_{j'} , \\
g_2' =& - g_2' - g_7' m_{j'} + g_8' m_j.
}
The equations above hold, however, only for constant couplings. Since, in general, $g_i$ and $g'_i$ will depend on $s_{12}$, $s_{13}$ and $s_{14}$, the relations in Eq.~\eqref{eq:coeffs_Majorana} will become more involved and will also depend on the particular way the coefficients depend on the Mandelstam variables.


\section{\label{sec:conclusions} Conclusions}

The phenomenon of neutrino oscillations is typically approached diagonalising the vacuum or matter Hamiltonian. In this paper, we have attacked the problem of how to describe neutrino oscillations using on-shell methods, {\it i.e.} without relying on fields, Lagrangians or mass matrices. As a first step, we have studied how to implement the notion of flavor from the point of view of scattering amplitudes. Then, we discussed how the PMNS matrix emerges in this framework. To the best of our knowledge, this is the first time that the flavor properties of amplitudes are discussed from a completely on-shell perspective.
To make contact with the usual field theoretical approach, we have also explicitly determined the UV origin (in operator language) of the different terms appearing in the 3-point amplitude in Sec.~\ref{sec:3_point_amp}, finding differences between the Dirac and Majorana neutrino cases. 
Finally, we proposed a prescription to compute the potential and the modified neutrino dispersion relations induced by the propagation over a charged lepton background. In particular, we were able to reproduce the known results for vector and scalar mediators, {\it i.e.} for those situations in which the medium spin does not count. We then used the formalism to compute the matter potential including all possible BSM contributions that can affect neutrino propagation by using the most general 4-point amplitude involving two neutrinos and two charged leptons.

\begin{acknowledgments}
We thank Renata Zukanovich Funchal for carefully reading a first version of the manuscript and for raising interesting points. GFSA acknowledges financial support from FAPESP under contracts 2019/04837-9
and 2020/08096-0. EB acknowledges financial support from FAPESP under contracts 2019/15149-6 and 2019/04837-9 and partial support from CNPq. GMS was partially supported by Coordenação de Aperfeiçoamento de Pessoal de Nível Superior (CAPES) and by Fundação de Amparo à Pesquisa de São Paulo (FAPESP) under contract 2020/14713-2.

\end{acknowledgments}

\appendix

\section{Conventions used}\label{app:conventions_spinor_helicity}

In this Appendix we summarise the conventions we use for the spinor variables. We use the bold notation of Ref. \cite{Arkani-Hamed:2017jhn} to denote massive spinors. They are explicitly given by
\be\label{eq:massive_helicity_up}
\bra{p^{I}} = 
\left(
\begin{array}{c|c}
\sqrt{E+|\bm{p}|} \,c & - \sqrt{E-|\bm{p}|}\, s \\
\sqrt{E+|\bm{p}|}\, s^* & \sqrt{E-|\bm{p}|} \,c
\end{array}\right),
\ee
\be
\sket{p^{ I}} = 
\left(\begin{array}{c|c}
(\sqrt{E-|\bm{p}|}\, s)^* & (\sqrt{E+|\bm{p}|} \,c)^* \\
- (\sqrt{E-|\bm{p}|}\, c)^* & (\sqrt{E+|\bm{p}|}\, s^*)^*
\end{array}\right) ,
\ee
with $I$ the $SU(2)$ little-group index, $c = \cos(\theta/2)$, $s = \sin(\theta/2) e^{i\phi}$ and $\sqrt{E\pm |\bm{p}|}$ a complex number. The first column refers to $I=1$, while the second corresponds to $I=2$. The spinors $\ket{p^I}$ and $\sbra{p^I}$ can be obtained by simply taking the complex conjugate. The helicity variables so defined satisfy the massive Weyl equations
\be\label{eq:Weyl1}
p \ket{p^I} = M \sket{p^I},\qquad p \sket{p^I} = M^\dag  \ket{ p^I},
\ee
\be\label{eq:Weyl2}
\sbra{p^I} p = - M^\dag \bra{p^I}, \qquad \bra{p^I} p = - M \sbra{p^I} ,
\ee
where the complex number $M$ is defined as $M M^\dag = m^2$, with $m$ the physical mass of the particle. Momentum bi-spinors are defined as
\be
p^{\dot{\alpha} \alpha} \equiv \epsilon_{IJ} \sket{p^I} \bra{p^J}, ~~~ p_{\alpha\dot{\alpha}} \equiv - \epsilon_{IJ} \ket{p^I} \sbra{p^J},
\ee
where
\be
   \epsilon_{IJ} = \begin{pmatrix}
        0 & -1 \\
        1 & 0 
    \end{pmatrix}.
\ee
The following identity involving two spinor variables of the same type have been used in the computation of the matter potential:
\be
\ket{p^I}_\alpha\bra{p_I}^\beta=M\delta_\alpha^\beta, ~~~~\sket{p^I}^{\dot{\alpha}}\sbra{p_I}_{\dot{\beta}}=-M^\dagger\delta^{\dot{\alpha}}_{\dot{\beta}},
\ee
while momentum conservation can be written as
\be
    \sum_i \ket{p_i^I} \sbra{p_{iI}} = 0.
\ee

Finally, we use the standard conventions for spinor variables with negative momenta:
\be
\bra{(-p)^I}=\bra{p^I}, \quad \sket{(-p)^I}=-\sket{p^I}.
\ee

\section{High energy basis}

The high energy limit for massive states can be studied by expanding the little group indexes in the following basis of SU(2) (we will follow the conventions from~\citep{Bachu:2019ehv})

\begin{align}\label{eq:high_energy_limit}
	\begin{split}	
		\ket{\lambda^I} &= \ket{\lambda}\,\xi^{-I} + 		
	 	\ket{\eta}\, \xi^{+I},\\
	 	|\tilde{\lambda}^{I}] &= |\tilde{\lambda}] \,\xi^{+I} + \sket{\tilde{\eta}}\, \xi^{-I},
	 \end{split}
\end{align}
where
\be
\xi^{+I}  = \begin{pmatrix} 0 \\ - 1 \end{pmatrix}, ~~~ \xi^{-I}  = \begin{pmatrix} 1 \\ 0 \end{pmatrix} .
\ee
The $SU(2)$ little group indices can be lowered with the antisymmetric tensor, giving
\be
\xi^{+}_I  = \begin{pmatrix} 1 \\ 0 \end{pmatrix},  ~~~ \xi^{-}_I  = \begin{pmatrix} 0 \\ 1 \end{pmatrix} .
\ee

The massless spinors in Eq.~\eqref{eq:high_energy_limit} can be read from Eq.~\eqref{eq:massive_helicity_up} as follows:
\al{
\ket{\lambda} & = \ket{p^1}, ~~ & \ket{\eta} & = \ket{p^2} , \\
|\tilde{\lambda}] & = \sket{p^2}, ~~ & \sket{\tilde{\eta}} & = \sket{p^1}  \\
}


With this identification we see that, in the high energy limit, terms containing $\ket{\eta}$ or $\sket{\tilde{\eta}}$, {\it i.e.} proportional to $\sqrt{E - p} \simeq m$ are subdominant with respect to terms containing $\ket{\lambda}$ and $|\tilde{\lambda}]$, justifying the results below Eq.~\eqref{eq:hel}. 

\section{Connection with the usual field theoretical derivation}\label{app:Dirac}

We now explicitly outline how the spinor helicity variables match into the more common notation in terms of Dirac spinors. Since we are discussing the interactions between fermions and vectors, it will be important to have the explicit expression for the polarisation vectors. For a massive spin-1 particle we have~\cite{Durieux:2019eor}
\be
\epsilon_\mu(p) = \frac{\asbraket{\bm{p}}{\sigma_\mu}{\bm{p}}}{\sqrt{2} m}\ ,
\ee
while for massless spin-1 particles we can write~\cite{Durieux:2019eor}
\be
\epsilon_\mu^{(+)}(p) = \frac{\asbraket{q}{\sigma_\mu}{p}}{\sqrt{2} \braket{pq}} , ~~~~~ \epsilon_\mu^{(-)}(p) = \frac{\asbraket{p}{\sigma_\mu}{q}}{\sqrt{2} \sbraket{pq}} ,
\ee
where $q$ is an arbitrary reference momentum. The spinor wave functions read~\cite{Dreiner:2008tw}
\al{
u^I(p) & = \begin{pmatrix} \ket{p^I} \\ \sket{p^I} \end{pmatrix}, & \bar{u}^I(p) = \begin{pmatrix} \bra{p^I}, & \sbra{p^I} \end{pmatrix}, \\
v^I(p) &=  \begin{pmatrix} \ket{p^I} \\- \sket{p^I} \end{pmatrix}, & \bar{v}^I(p) = \begin{pmatrix} - \bra{p^I}, & \sbra{p^I} \end{pmatrix}, 
}
and we are using the Weyl representation of the Dirac matrices. 

For the monopole interactions of massive particles we obtain
\al{
\bar{v}_2 \gamma^\mu P_L u_1 \epsilon_\mu(p_3) & = \sabraket{\bm{2}}{\bar{\sigma}^\mu}{\bm{1}} \frac{\asbraket{\bm{3}}{\sigma_\mu}{\bm{3}}}{\sqrt{2} m_3} = -\frac{ \sqrt{2} \braket{\bm{13}} \sbraket{\bm{23}} }{m_3} ,\\
\bar{v}_2 \gamma^\mu P_R u_1 \epsilon_\mu(p_3) & = - \sabraket{\bm{1}}{\bar{\sigma}^\mu}{\bm{2}} \frac{\asbraket{\bm{3}}{\sigma_\mu}{\bm{3}}}{\sqrt{2} m_3} =  \frac{\sqrt{2} \braket{\bm{23}} \sbraket{\bm{13}} }{m_3} ,
}
where we have used the identity $\bar{\sigma}^{\mu \dot{\beta}\beta} \sigma_{\mu \alpha \dot{\alpha}} = 2 \delta_\alpha^\beta \delta_{\dot{\alpha}}^{\dot{\beta}} $. 
As observed around Eq.~\eqref{eq:massless_limit}, choosing the coefficient to be inversely proportional to the vector mass ensures the consistency of the massless limit. We see here that the same factor appears from the direct computation of the amplitude. 

For dipole interactions we instead obtain
\al{
 \bar{v}_2 \sigma^{\mu\nu} P_L  u_1 \left(p_{3\mu} \epsilon_\nu(3) - p_{3\nu} \epsilon_\mu(3) \right) 
&=  -2\sqrt{2} \braket{\bm{13}} \braket{\bm{23}} , \\
\bar{v}_2 \sigma^{\mu\nu} P_R  u_1 \left(p_{3\mu} \epsilon_\nu(3) - p_{3\nu} \epsilon_\mu(3) \right) 
&=  -2\sqrt{2} \sbraket{\bm{13}} \sbraket{\bm{23}} . \\
}
\vspace{1em}

\bibliographystyle{JHEP2}
\bibliography{on-shell_bib}

\end{document}